\documentclass[aps,twocolumn, floatfix, prx,longbibliography]{revtex4-1}
\usepackage{graphicx}
\DeclareGraphicsExtensions{.pdf}
\usepackage{amsmath,amssymb,bbold,bm,color}
\usepackage{float,hyperref}
\usepackage{epstopdf}
\usepackage{titlesec}
\hypersetup{colorlinks=true,urlcolor=blue}
\usepackage[percent]{overpic}

\begin{document}

\title{Altermagnetism in modified Lieb lattice Hubbard model}
\author{Nitin Kaushal}
%\affiliation{Department of Physics and Astronomy, The University of 
%Tennessee, Knoxville, Tennessee 37996, USA}
\affiliation{Department of Physics and Astronomy and Quantum Matter Institute, University of British Columbia, Vancouver, British Columbia BC V6T 1Z4, Canada}
\author{Marcel Franz}
%\affiliation{Department of Physics and Astronomy, The University of 
%Tennessee, Knoxville, Tennessee 37996, USA}
\affiliation{Department of Physics and Astronomy and Quantum Matter Institute, University of British Columbia, Vancouver, British Columbia BC V6T 1Z4, Canada}
%\date{June 2024}
\date{\today}

\begin{abstract}
 We study the emergence of altermagnetism from repulsive interactions
for electrons on the Lieb lattice as a model of quasi-2D
oxychalcogenides with the so-called ``anti-CuO$_{2}$'' lattice structure. 
 A comprehensive study of the Lieb lattice Hubbard model, using unrestricted Hartree-Fock and exact diagonalization techniques,
 establishes the presence of spin-${1\over 2}$ altermagnetic Mott insulating ground state for average electron densities of 2 and 4 per unit cell. 
 %Our exact dynamical spin structure factor calculations on an effective spin model confirm the presence of chiral symmetry breaking in collective spin excitations in the altermagnetic Mott state. 
Both phases show the characteristic spin splitting in the electron bands as well as in the magnon bands, as indicated by solutions of an effective spin-${1\over 2}$ Heisenberg model that we construct.   We also provide evidence for altermagnetic metal formation in the electron- and hole-doped Mott state, giving rise to Fermi surfaces with $d_{x^{2}-y^{2}}$-wave spin splitting and quasi-one-dimensional characteristics. 
\end{abstract}
\maketitle

\textit{Introduction.--}
In recent years, the altermagnetic state has been established as a
distinct class of collinear magnets that has zero net magnetization,
akin to conventional antiferromagnets, accompanied by broken Kramer's
degeneracy through non-relativistic spin splitting in single-particle
electron bands~\cite{Hayami01,Smejkal03,Yuan01,Mazin01}. The above contrasting features are simultaneously present when the atomic positions with opposite spins are related by spatial rotation or mirror reflection, but not by spatial inversion or translation~\cite{Smejkal01,Smejkal02}. The large spin-splitting in bands, possibly of the order of eV, provides a promising avenue for future spintronics applications. Moreover, many theoretical proposals for interesting topological phenomena in altermagnets have also been advanced~\cite{YLi01,Zhu01,Heung01}.

The above findings have stimulated a wave of interest in candidate
materials showing altermagnetism. Angle-resolved photoemission (ARPES)
measurements have recently confirmed $g$-wave altermagnetic band
splitting in three-dimensional candidate materials
CrSb~\cite{Ding01,Lu01,CLi01} and
MnTe~\cite{JKrempasky01,Lee01,Osumi01}. However, the status of
$d$-wave altermagnetism in RuO$_2$ remains
unresolved~\cite{Liu01,Kessler01}. Multiple theoretical proposals for
two-dimensional (2D) altermagnetic materials, based on {\rm ab-initio}
calculations have been also made~\cite{Mazin02}. Recently, a Mn-based
quasi-2D oxyselenide La$_2$O$_3$Mn$_2$Se$_2$, with collinear
antiferromagnetic ordering observed by neutron-diffraction studies,
was proposed to be a $d_{x^{2}-y^{2}}$-wave altermagnetic
insulator~\cite{Ni01,Wei01}. However, other materials belonging to the
same class, namely the Co- and Fe-based oxyselenides, have shown
non-collinear orthogonal spin
arrangements~\cite{Fuwa01,Freelon01,LZhao01}. In parallel, compelling
experimental evidence for room-temperature $d$-wave altermagnetism was
reported in other quasi-2D oxychalcogenide metals,
KV$_2$Se$_2$O~\cite{BJiang01} and
Rb$_{1-\delta}$V$_{2}$Te$_{2}$O~\cite{FZhang01}, using spin resolved
ARPES.
\begin{figure}[!t]
\hspace*{0.1cm}
\vspace*{0cm}
\includegraphics[width = 7.5cm]{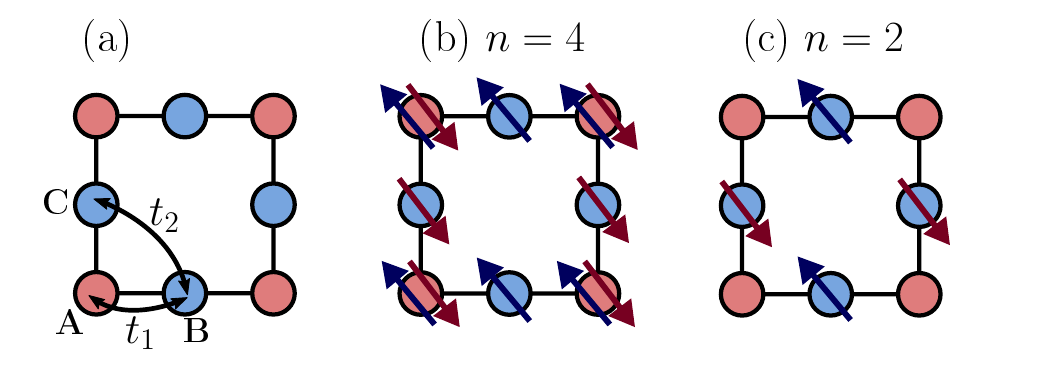}
\caption{Modified Lieb lattice used in the present work. In the anti-CuO$_2$ structure the A site is occupied by oxygen while B, C sites host a transition metal.  Panels (b) and (c) show the pictorial representation of the antiferromagnetic state for average electron density of 4 and 2, respectively.}
\label{fig1}
\end{figure}

The crucial characteristic underlying the emergence of  $d$-wave
altermagnetism in oxychalcogenides is their ``anti-CuO$_{2}$''
structure' of the $T_{2}$O layers consisting of transition metal
($T$) and oxygen (O) atoms in the Lieb lattice arrangement,  Fig.~\ref{fig1}(a). In these 
$T_{2}$O layers, a different crystallographic environment is provided
by the neighboring non-magnetic atoms (oxygen and chalcogen) to
the two transition metal atoms. Importantly, these two distinct
transition metal positions are connected by a $C_{4}$ rotation, but
not by inversion or lattice translation.  If one ignores the chalcogen
atoms that are present out of the $T_{2}$O planes then the minimal 2D
model capturing the above lattice symmetry is the Lieb lattice model
with 3 sublattices per unit cell, where  sublattice A corresponds to
oxygen atoms and sublattices B and C represent the transition metal
atoms, see Fig.~\ref{fig1}(a). 

A spin-fermion model on the Lieb lattice,
with sublattices B and C coupled to static
antiferromagnetically ordered magnetic moments, has been shown to
exhibit a $d_{x^2  -y^2}$-wave altermagnetic band
splitting~\cite{Brekke01,Antonenko01}.
In this Letter, instead of presuming the magnetic order, we
undertake a comprehensive and unbiased study of magnetism emerging
from Coulomb interactions in the Lieb lattice model. This allows us to
uncover the conditions under which the altermagnetic ground state is
stabilized over other competing magnetic states. It should be noted that
a variety of other lattices with Hubbard interaction have been
proposed to stabilize
altermagnets~\cite{Maier01,Leeb01,Das01,Roig01,Ferrari01,Re01,Giuli01}. However,
unbiased studies establishing altermagnet as a ground state of an
interacting Hamiltonian are lacking, especially on the realistic
lattice geometries. 

Here we study the Lieb lattice with the local
Hubbard interaction present only on sublattices B and C containing the
magnetic atoms,  while maintaining sublattice A
non-magnetic. We use unrestricted Hartree-Fock approximation, without
assuming any ansatz magnetic order, see Supplementary Material (SM) ~\cite{Supple} for details,
to map out the phase diagram as a function of the next-nearest neighbor
hopping and the Hubbard interaction strength for various electron
densities. Moreover, to
confirm the stability of the Hartree-Fock results against quantum
fluctuations, we perform an exact diagonalization (ED) study
of the interacting model.  We also discuss the spin dynamics of the altermagnetic
Mott state, using the effective spin-${1\over2}$ model. Finally, we
investigate the possibility of stabilizing the altermagnetic metal by
doping the altermagnetic Mott insulator away from integer fillings. 

\textit{Model.---}
We study the two-dimensional Lieb lattice Hubbard model, 
\begin{multline}\label{Hamil}
%    \hspace{9.0em}
H = t_{1}\sum_{\langle i,j\rangle,\sigma}c_{i,\sigma}^{\dagger}c_{j,\sigma} + 
    t_{2}\sum_{\langle \langle i,j\rangle\rangle,\sigma}c_{i,\sigma}^{\dagger}c_{j,\sigma} \\ 
    + U\sum_{i\in \{B,C\}}n_{i,\uparrow}n_{i,\downarrow} + \epsilon_{A}\sum_{i\in A}n_{i} - \mu\sum_{i}n_{i},
    \hspace{1.8em}
\end{multline}
% see Fig.\ref{fig1}(a)
where $c_{i,\sigma}^{\dagger}$ creates an electron at site $i$ with
spin $\sigma$,  $n_{i,\sigma}=c_{i,\sigma}^{\dagger}c_{i,\sigma}$
is the density operator and $n_i=\sum_\sigma n_{i,\sigma}$. The site index $i$ labels the
unit cell and the sublattice degree of freedom. Parameters $t_{1}$ and
$t_{2}$ represent the nearest and next-nearest neighbor hopping
amplitudes, respectively. The local Coulomb repulsion is parameterized
by $U$, which is present only on sublattices B and C in
Fig.\ref{fig1}(a). Parameter $\epsilon_{A}$ denotes the onsite
energy of sublattice A. We tune the chemical potential $\mu$ to attain
the targeted average electronic density per unit cell
$n=\frac{1}{L_{x}L_{y}}\sum_{i}\langle n_{i}\rangle$, where $L_{x}
(L_{y})$ is the number of unit cells in $x (y)$ direction. In all our
calculations we fix $t_{1}=-1.0$ and keep $|t_{1}|$ as a unit of energy. 

\begin{figure}[!t]
\hspace*{-0.5cm}
\vspace*{0cm}
\begin{overpic}[width=1.1\columnwidth]{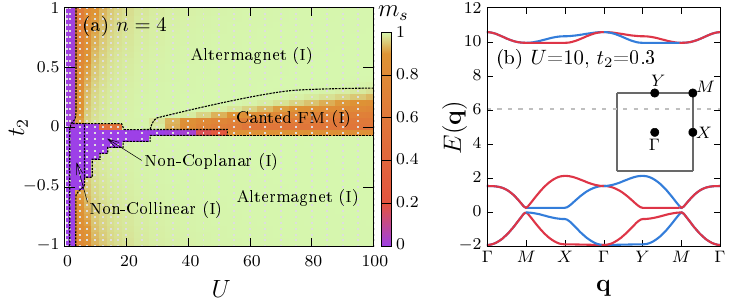}
\end{overpic}
\caption{Phase diagram of the Hubbard model Eq.\ \eqref{Hamil}  for $\epsilon_{A}=0$ and  $n=4$ obtained by Hartree-Fock approximation, panel (a). The violet colored region near $U=0$ comprises paramagnetic metal; ``I''  stands for Insulator.  Panel (b) shows the band structure of the altermagnetic insulator for parameters $U=10$, $t_{2}=0.3$, and $n=4$ illustrating the spin-splitting in spin up (blue) and down (red) bands. The dashed line represent the chemical potential. }
\label{fig2}
\end{figure}

\textit{Altermagnetic Mott insulators at $n=2,4$.---}
We expect antiferromagnetic (AF) Mott insulators to form at these fillings which may be understood by considering some limiting cases. For $ n=4$, in the limit $U\gg(|t_{1(2)}|,\epsilon_{A})$, the double occupancy on sublattices B and C must be suppressed leading to $n_{A}=2$ and $n_{B}= n_{C}=1$ configuration, depicted in Fig.~\ref{fig1}(b). On the other hand, for $n=2$, in the limit $\epsilon_{A}\gg t_{1}$, sublattice A should be empty ($n_{A}=0$) which again favors $n_{B}= n_{C}=1$; the corresponding configuration is shown in Fig.~\ref{fig1}(c). In both cases, in the limit of large $U$, the antiferromagnetic correlations between half-filled B and C sites are expected to form due to Anderson superexchange $J\sim  t_{2}^{2}/U$. We shall see that AF Mott insulators indeed emerge at $n=2,4$ even away from these limits. Furthermore, they exhibit spin-split electron bands and should be therefore regarded as altermagnetic Mott insulators.

In a conventional Lieb lattice Hubbard model, the $n =4$ and $n=2$ cases are related by a particle-hole symmetry \cite{Gouveia01}. The Hamiltonian \eqref{Hamil} used in the present work, for $t_{2}=\epsilon_{A}=U=0$,  is also invariant under a particle-hole transformation $c_{j\alpha\sigma}\rightarrow v_{\alpha} c_{j\alpha\sigma}^{\dagger}$, where $j,\alpha$, and $\sigma$ represent the unit cell, sublattice, and spin index, respectively and $v_{\alpha}$=$\{-1,1,1\}$ for $\alpha$=$\{A,B,C\}$. On the other hand for $t_{2}=\epsilon_{A}=0$ and finite $U$, the particle-hole transformation gives $H \rightarrow H + U(2L_{x}L_{y}-N) + UN_{A}$, where $N$=$\sum_{j,\alpha,\sigma}n_{j,\alpha,\sigma}$ and $N_{A}$=$\sum_{j,\sigma}n_{j,A,\sigma}$. Because $[H,N_{A}]\ne 0$, the modified Lieb lattice Hamiltonian \eqref{Hamil}  does not respect the particle-hole symmetry for $U\ne 0$ and hence requires separate investigations for $n=2,4$ cases.
\begin{figure}[!b]
\hspace*{0.0cm}
\vspace*{0cm}
\begin{overpic}[width=1.0\columnwidth]{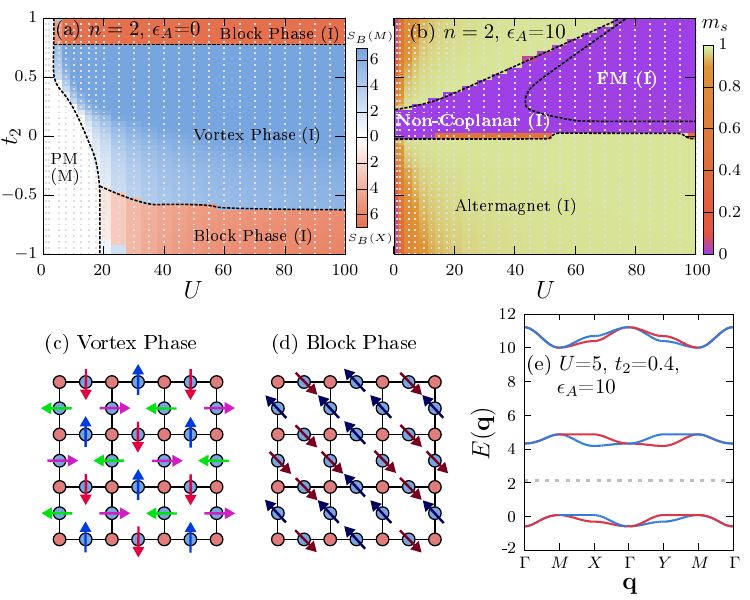}
\end{overpic}
\caption{Results for $n=2$. Panels (a) and (b) show phase diagrams for $\epsilon_{A}=0$ and $\epsilon_{A}=10$, respectively. Real-space pictorial representation of the vortex (block) state is depicted in panel c (d). Panel (e) shows the band structure of the altermagnetic insulator, choosing the representative parameter point of $U=5$, $t_{2}=0.4$, $\epsilon_{A}=10$.}
\label{fig3}
\end{figure}

First, we discuss the results for filling $n=4$ obtained by unrestricted Hartree-Fock calculations for the system size of $L_{x}$$\times$$L_{y}$=$12$$\times$$12$. In Fig.~\ref{fig2}(a), we show the $t_{2}$ vs $U$ phase diagram for fixed $\epsilon_{A}=0$.   We find that in a large part of the phase diagram, spanning weak to strong $U$, the Hartree-Fock ground state shows a collinear AF order in spins on sublattices B and C. We use staggered and total magnetization
\begin{equation}
  m_{s/t}=\frac{1}{L_{x}L_{y}}\big|\sum_{i} \langle{\mathbf{S}}_{i,B}\mp {\mathbf{S}}_{i,C}\rangle\big|
\end{equation}
 to quantify the strength of collinear AF and FM order in the phase diagram.  Fig.~\ref{fig2}(b) confirms the expected  $d$-wave spin splitting in the single-electron bands in the altermagnetic region, with a gap at the chemical potential. The splitting between the spin-up (blue) and spin-down (red) bands is present except for the nodal lines $k_{x}=\pm k_{y}$ indicating a $d_{x^{2} - y^{2}}$ symmetry. The spin degeneracy on these nodal lines is protected by the $C_{4}$ rotational symmetry of the lattice. We also notice a canted ferromagnetic (FM) state, in the large $U$ region for small $t_{2}>0$, where the canting angle between ${\mathbf{S}}_{i,B}$ and ${\mathbf{S}}_{i,C}$ continuously increases to $\pi$ as $t_{2}$ is increased and smoothly transits into the altermagnetic state. The total magnetization $m_{t}$ is zero everywhere in the phase diagram except in the canted FM region. For $t_{2}<0$ the ground state is magnetically frustrated up to intermediate $U$ values and shows non-collinear vortex spin order (see Fig.~\ref{fig2}(c)) and non-coplanar magnetism. We provide a more detailed discussion of the non-coplanar state in SM ~\cite{Supple}.

 Fig.~\ref{fig3}(a) shows the $t_{2}$ vs $U$ phase diagram for $n=2$ and fixed $\epsilon_{A}=0$ as above. Unlike the $n=4$ case, we find mainly two magnetic insulators: the vortex state and the block state. The non-collinear vortex state is stable in large part of the phase diagram, for $-0.65\lesssim  t_{2}\lesssim 0.80$. In this state, ${\mathbf{S}}_{i,B}$ and ${\mathbf{S}}_{i,C}$ are aligned orthogonally making vortices with staggered chirality as shown in Fig.~\ref{fig3}(c). Interestingly, this vortex state has been observed in neutron diffraction experiments on a Co based oxychalcogenide La$_{2}$Co$_2$O$_3$Se$_{2}$~\cite{Fuwa01} which has anti-CuO$_{2}$ structure in Co$_{2}$O layers resembling the Lieb lattice used in our study. This state shows a peak in the sublattice resolved spin structure factor
\begin{equation}
  S_{\alpha}({\bf q})={1\over L_{x}L_{y}}\sum_{i,j}{\bf{S}}_{\alpha,i}\cdot {\bf{S}}_{\alpha,j}e^{\iota({\bf{r}}_{i} - {\bf{r}}_{j})\cdot {\bf{q}}},
\end{equation}
  at momentum ${\bf q}=M$ for both $\alpha=B$ and C. For large $|t_{2}|$ AF block state is stabilized with $2\times 2$ ferromagnetic blocks, see Fig.~\ref{fig3}(d). For the block state, $S_{B(C)}({\bf q})$ peaks at ${\bf q}=X(Y)$ reflecting the AF and FM correlations in $x(y)$ and $y(x)$ directions, respectively. In Fig.~\ref{fig3}(a), we used $S_{B}(M)$ and $S_{B}(X)$ to differentiate between the regions with vortex and block ordering. The apparent differences between the phase diagrams of Fig.~\ref{fig2}(a) and Fig.~\ref{fig3}(a) are related to broken particle-hole symmetry discussed earlier.

Fixing $\epsilon_{A}=10$ leads to nearly half-filled $B$ and $C$ sublattices and stabilizes altermagnetic insulating state in a large part of the phase diagram  Fig.~\ref{fig3}(b), especially when $t_{2}<0$. For $t_{2}>0$, we find a saturated ferromagnet in the large $U$ region and a non-coplanar magnetic order for weak and intermediate values of $U$, for details see supplementary~\cite{Supple}. In Fig.\ \ref{fig3}(e), we show the band structure for the altermagnetic phase depicting once again the $d_{x^{2}-y^{2}}$-wave spin splitting. We note that for both $n=4$ and 2, the next nearest neighbor hopping is required to stabilize the altermagnetic state in the large range of $U$. 

\begin{figure}[!t]
\hspace*{0.0cm}
\vspace*{0cm}
\begin{overpic}[width=1.0\columnwidth]{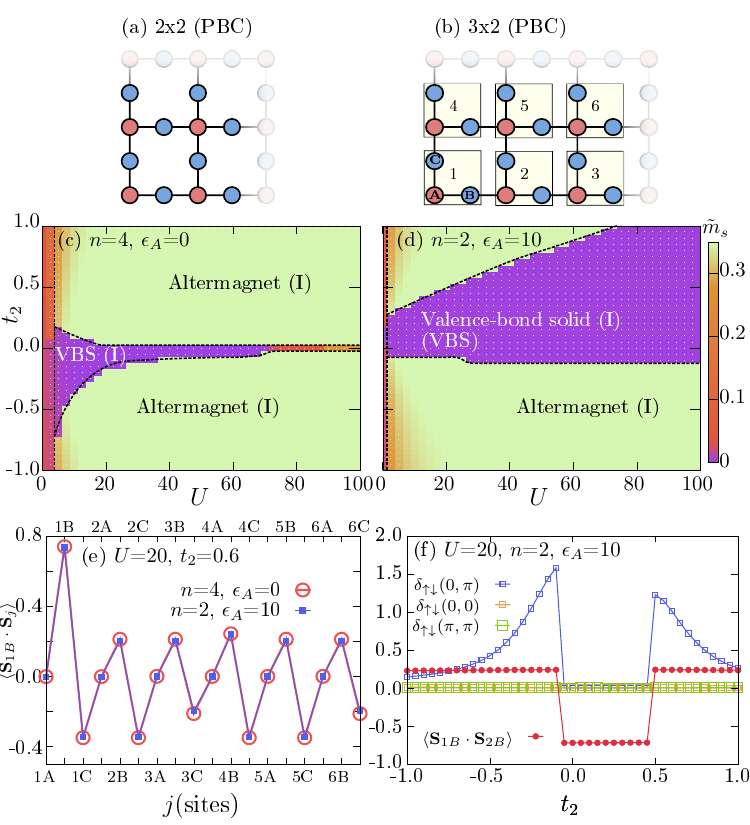}
\end{overpic}
\caption{ED results for 12-site ($2\times2$) and 18-site ($3\times2$) clusters with PBC; the corresponding lattices are shown in panels (a,b). Phase diagrams for $n=4$ and $n=2$ are shown in panels (c) and (d), respectively, using the 12-site cluster. The spin-spin correlation w.r.t.\ site $1B$, using the 18-site cluster, are shown in panel (e), choosing two parameter points representing the altermagnetic state. Panel (f) shows the evolution of spin splitting in the spectral function $\delta_{\uparrow\downarrow}({\mathbf{k}})$ and spin-spin correlation between site 1B and 2B, as a function of $t_{2}$.}
\label{fig4}
\end{figure}

\textit{Exact diagonalization.---} To confirm the validity of our mean-field results  we  performed ED studies of the Hubbard model Eq.\  \eqref{Hamil} on small  clusters with 12 sites  and periodic boundary conditions (PBC) which corresponds to $2\times 2$ in the unit cell nomenclature of Fig.~\ref{fig4}(a).
We used staggered magnetization, calculated using correlations as $\tilde{m}_{s}={(L_{x}L_{y})^{-2}}\sum_{i,j}\langle{\mathbf{S}}_{iB}\cdot ({\mathbf{S}}_{jB} - {\mathbf{S}}_{jC} )\rangle$ to identify the region with collinear AF order in the phase diagrams. We explored the $t_{2}$ vs $U$ phase diagrams, as in our Hartree-Fock calculations, for ($ n=4$, $\epsilon_{A}=0$) and ($ n =2$, $\epsilon_{A}=10$) cases shown in Figs.~\ref{fig4}(c) and \ref{fig4}(d), respectively. We found robust staggered magnetization in all the regions where Hartree-Fock predicted the altermagnetic ground states. We also noticed that instead of the canted FM state in $n =4$ phase diagram at large $U$, exact results indicate staggered magnetization. Moreover, for all the parameter values of Fig.~\ref{fig4}(c) and (d), the ground state has a total $S=\sum_{i,j}{\langle \mathbf{S}_{i}\cdot\mathbf{S}_{j}\rangle}=0$ suggesting the absence of any FM states contrary to mean-field calculations. For the regions in the Hartree-Fock phase diagrams Fig.~\ref{fig2}(a) and Fig.~\ref{fig3}(b) where we found frustrated non-collinear, non-coplanar, or saturated FM order, the exact results suggest instead a valence bond solid (VBS) state (see Fig.~\ref{fig4}(a)). In the VBS, for $2\times 2$ system, we found 
that electrons on the sublattice B (A) are present in a singlet state along $x (y)$ direction with $\langle {\bf S}_{i}\cdot {\bf S}_{j}\rangle \approx -0.75$. We note, however, that confirming  the stability of VBS against a spin liquid or other AF ordered states in a larger system would require calculations beyond the scope of the present work.

We also investigated the $3\times 2$ (18-site cluster) with PBC, see Fig.~\ref{fig4}(b). The observed spin-spin correlations with respect to the site `1B' in Fig.~\ref{fig4}(e) for two parameter points, again suggest robust AF correlations between sublattices B and C. To investigate altermagnetism, we define momentum-resolved spin splitting  $\delta_{\uparrow\downarrow}(\mathbf{k})=\int_{}^{} d\omega (A^{h}_{\uparrow}({\mathbf{k},\omega})-A^{h}_{\downarrow}({\mathbf{k},\omega}))^{2}$, where $A_{\sigma}^{h}({\bf k},\omega)$ is the hole spectral function (for details see ~\cite{Supple}). It should be noted that in exact calculations on small clusters, the time-reversal symmetry is not broken spontaneously so we add a small Zeeman field of $0.05$ on a corner site `1B' in our calculations while calculating $\delta_{\uparrow\downarrow}(\mathbf{k})$. In Fig.~\ref{fig4}(f), we show the results for $2\times 2$ system because this geometry does not break the rotational symmetry. We fix $n=2$, $\epsilon_{A}=10$,   $U=20$ and tune $t_{2}$ so that the system transits through multiple phases. Spin splitting measured by $\delta_{\uparrow\downarrow}(\mathbf{k})$  at ${\bf k}=(0,\pi)$ is non-zero in the region with collinear AF correlations, whereas it vanishes for ${\bf k}=(0,0)$ and $(\pi,\pi)$. Moreover, VBS state does not show any altermagnetic splitting. These findings are consistent with our Hartree-Fock results and support the altermagnetic Mott insulator ground state. 
\begin{figure}[!t]
\hspace*{0.0cm}
\vspace*{0cm}
\begin{overpic}[width=1.0\columnwidth]{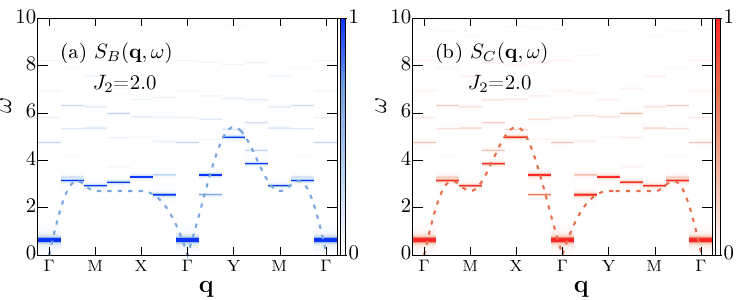}
\end{overpic}
\caption{Dynamical spin structure factors of the effective spin model for sublattice B and C calculated using ED. Dashed lines correspond to the linear spin wave theory results. }
\label{fig5}
\end{figure}

\textit{Effective Spin model.---}
In the altermagnetic Mott insulator we can use an effective spin-${1\over 2}$ model to investigate the spin excitations. Only the half-filled sublattices B and C are relevant for the spin model because the spin moments on empty or doubly occupied sublattice A are quenched, leading to a realization of spin-${1\over 2}$ Heisenberg model on a checkerboard lattice. The model Hamiltonian can be written as
\begin{multline}\label{HamilEff}
H_{\textrm{eff}}=J_{2}\sum_{\langle ij \rangle} {\mathbf {S}}_{i}\cdot {\mathbf {S}}_{j} + J_{1}\sum_{\langle \langle ij \rangle \rangle}  {\mathbf {S}}_{i}\cdot {\mathbf {S}}_{j},
\hspace{2.8em}
\end{multline}
where $J_{2}$ and $J_{1}$ denote the first and second neighbor exchange couplings on the checkerboard pattern of plaquettes, respectively. The AF exchange $J_{1}$ is predominantly driven by 4th order process via hopping $t_{1}$, whereas $J_{2}$ has considerable contributions also from 2nd order process via $t_{2}$ and 3rd order via combination of $t_{2}$ and $t_{1}$ ~\cite{Supple}. 

We investigate the spin model by treating $J_{2}$ as a free parameter and fixing $J_{1}=1$. We solve a 32-site cluster with PBC using exact diagonalization, while targeting the total momentum ${\bf K}$ and total $S_{z}$. The ground state is always present in total momentum ${\bf K}=(0,0)$ and $S_{z}=0$ sector, with total $S=\sum_{ij}{\bf S}_{i}\cdot{\bf S}_{j}=0$. We find a first order phase transition near $J_{2}=1.0$, from sliding Luttinger spin liquid at small $J_2$ to an altermagnet at large $J_2$~\cite{Sindzingre01,YHChan01,SMoukari01,RFBishop01,Supple}.  Fig.~\ref{fig5} shows a comparison of the sublattice-resolved spin spectrum of Hamiltonian Eq.\ \eqref{HamilEff} obtained using linear spin wave theory (LSWT) and  the dynamical spin structure factor  $S_{\alpha}({\bf q},\omega)$ extracted from ED, both in the altermagnetic phase. The Goldstone-like magnon for $S_{B(C)}({\bf q},\omega)$ emerging from the momentum $\Gamma$ is consistent with having long-range correlations within sublattice B (C). Meanwhile $S_{B}({\bf q},\omega)$ and $S_{C}({\bf q},\omega)$ are different but  related by a $90^{\circ}$ rotation, showing the characteristic altermagnetic splitting which could be observed experimentally using inelastic neutron scattering.

\begin{figure}[!t]
\hspace*{0.0cm}
\vspace*{0cm}
\begin{overpic}[width=1.0\columnwidth]{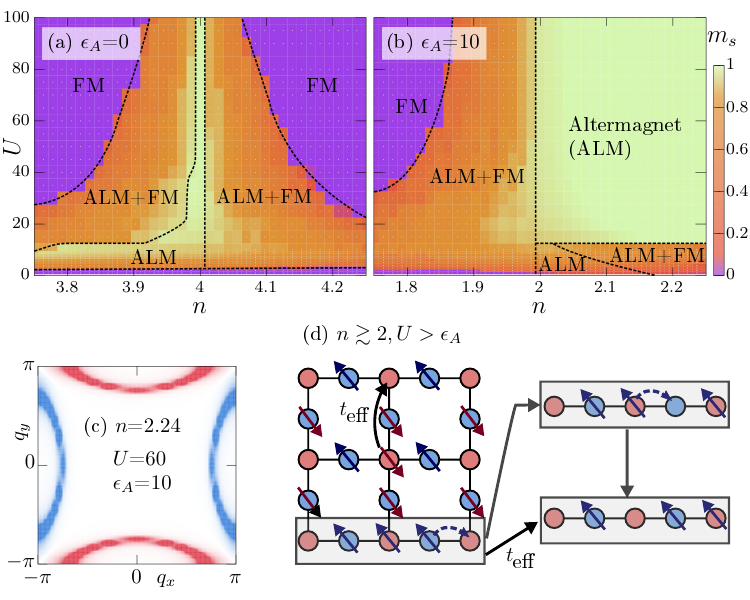}
\end{overpic}
\caption{Panels (a) and (b) show phase diagrams around the fillings 4 and 2, respectively. Panels (c) depicts the spin-resolved Fermi surface for the parameters described, with the blue (red) color corresponding to spin up (down) bands. Panel (d) depict the quasi-1D kinetic motion of the doped electrons near the filling 2. $t_{2}$ is fixed to -1.0 for all of the above results.}
\label{fig6}
\end{figure}

\textit{Altermagnetic metal by doping.---} Having established the existence of altermagnetic Mott insulators in the Lieb lattice model an interesting question arrises: Is it possible to stabilize an altermagnetic metal by doping away from integer fillings? Our unrestricted Hartree-Fock approach answers this question in the affirmative. As seen in Fig.~\ref{fig6}(a) the altermagnetic (ALM) metal emerges at weak hole doping of the $n=4$ Mott insulator and is stable over an extended hole density at small $U$. At larger $U$ and for electron doping our results indicate inhomogeneous states comprised of ALM and FM regions interpolating towards pure FM phases.  Similar inhomogeneous states have been discussed in doped square lattice Hubbard models near the half-filled Mott state~\cite{PIgoshev01,JXu01}. In general, for 2D bi-partite lattices, the inhomogeneous states are expected on doping the AF insulator because the doped fermions favor ferromagnetic background in their spatial local environment~\cite{RSamajdar01,RSamajdar02}. A more detailed discussion of the ALM+FM state is given in SM \cite{Supple}. 

Remarkably, the altermagnetic metal is stable in a large part of the phase diagram reached by electron-doping the $n=2$ insulator for  $U>10$, Fig.~\ref{fig6}(b). In this region the doped electrons are mostly confined to the sublattice A because of the cost of double occupancy in sublattices B and C. Quasi-1D electron pockets emerge near $X\rightarrow M$ (spin-up) and  $Y\rightarrow M$ (spin-down) momentum showing the characteristic ALM spin splitting, see Fig.~\ref{fig6}(c).
This quasi-1D behavior arises from the restricted motion of doped spin-up (down) electrons on sublattice A via a second-order process along the $x$ ($y$) direction, depicted in Fig.~\ref{fig6}(d), with effective hopping $t_{\rm eff}\approx t_{1}^{2}/\epsilon_{A}$. We also found a small region of altermagnetic metal for $U<\epsilon_{A}$ and $n>2$ which resembles the $n<4$ ALM metal (see~\cite{Supple} for details). 

Finally, it should be noted that the Hartree-Fock approximation tends to overestimate the stability of itinerant FM phases~\cite{THanisch01,VBach01,BShastry01}. Hence, it is likely that the ALM metal can be stable over a larger part of the phase diagram than indicated in Fig.~\ref{fig6}(a,b). 

\textit{Conclusions.---} Our unbiased study of the  Lieb lattice Hubbard model adapted to the  anti-CuO$_{2}$ structure of oxychalcogenides establishes the presence of altermagnetic Mott insulators for electron densities $n=2,4$ in a broad range of the Coulomb interaction strength, in the presence of next-nearest neighbor hopping. The electron bands show the characteristic $d_{{x}^{2}-y^{2}}$-wave spin splitting expected of altermagnets which is observable using angle-resolve photoemission. An effective spin model on the checkerboard lattice likewise shows chiral splitting between spin excitations in the sublattice resolved dynamical spin-structure factor which matches results based on the linear spin wave theory and is measurable using inelastic neutron scattering. Doping away from integer fillings can stabilize metallic altermagnets which are of interest for spintronic applications. 

Quasi-2D materials with anti-CuO$_{2}$ structure and $d^{1}$ electronic configuration~\cite{TYajima01,XYan01} are natural candidates to realize the phases discussed in the present work. On the conceptual level our work shows how altermagnetism can emerge from a simple interacting model in a realistic geometry. It also provides a starting point for various future studies, allowing exploration of correlation-driven superconductivity in doped altermagnetic Mott insulators~\cite{Dagotto01}, and of topological superconductivity predicted to occur in altermagnetic metals \cite{Zhu01, Heung01}. Adding multiple $d$-orbitals is also an appealing direction which may help to understand complex magnetism in the family of quasi-2D oxychalcogenides recently identified as $d$-wave altermagnet candidates. 

{\em Note added --} While completing this work, another work appeared where altermagnetism is studied on the standard Lieb lattice Hubbard model~\cite{MDurrnagel01}. The crucial difference in our model is the absence of Coulomb repulsion on sublattice A.

{\em Acknowledgments --} The authors are indebted to N.\ Heinsdorf, R. Valenti, A. Nocera, and A. Patri
for stimulating discussions and correspondence. The work
was supported by NSERC, CIFAR and the Canada First Research Excellence Fund, Quantum Materials and Future Technologies Program.

%-----------------------------------------------------------------

%-----------------------------------------------------------------
\newpage
\onecolumngrid{
\raggedbottom
\begin{center}
{\bf \uppercase{Supplementary Information}} for\\
\vspace{10pt}
{\bf \large Altermagnetism in modified Lieb lattice Hubbard model}\\
\vspace{10pt}
by N. Kaushal and M. Franz
\date{\today}
\end{center}
%\date{\today}

%%%%%%%%%% Prefix a "S" to all equations, figures, tables and reset the counter %%%%%%%%%%
\setcounter{equation}{0}
\setcounter{figure}{0}
\setcounter{table}{0}
\setcounter{page}{1}
\makeatletter
\renewcommand{\theequation}{S\arabic{equation}}
\renewcommand{\thefigure}{S\arabic{figure}}
%\renewcommand{\bibnumfmt}[1]{[S#1]}
%\renewcommand{\citenumfont}[1]{S#1}
%%%%%%%%%% Prefix a "S" to all equations, figures, tables and reset the counter %%%%%%%%%%

%\maketitle
\section{Details of unrestricted Hartree-Fock technique}
In this section, we discuss the details of the unrestricted Hartree-Fock technique used in the present work.  Firstly, we perform the Hartree-Fock decomposition on the interaction part of the Hamiltonian, $U\sum_{j,\alpha \in \{B,C\}} n_{j \alpha \uparrow} n_{j \alpha \downarrow}$, where $j$ and $\alpha$ correspond to the unit cell and the sublattice index, respectively,
\begin{equation}\label{HFRealS}
H_{\rm int}^{HF} = U\sum_{\substack {j, \sigma \\ \alpha \in \{B,C\} }} \langle n_{j \alpha \bar{\sigma}}\rangle  n_{j \alpha {\sigma}} - U\sum_{\substack{j \\ \alpha \in \{B,C\} } } \left(\langle S_{j \alpha}^{+}\rangle S_{j \alpha}^{-}  + {\rm h.c.}\right).
%\hspace{22.8em}
\end{equation}
Furthermore, we assume that the mean-field result breaks the translational symmetry and emergent magnetic unit cell has the size of $l^{m}_{x}\times l^{m}_{y}$. The above assumption can be imposed on the local order parameters using $O_{l \beta \alpha} = O_{\beta \alpha}$, where $l=(l_{x},l_{y})$ is the magnetic unit cell index, $\beta$ is the internal unit-cell index in the $l$th magnetic unit cell, and $\alpha$ is the sublattice index as before. We write Eq.~\eqref{HFRealS} in hybrid real/momentum space representation using $c_{l \beta \alpha \sigma}^{\dagger}=\frac{1}{\sqrt{l^{m}_{x}l^{m}_{y}}} \sum_{{\bf k}}e^{\iota {\bf k}\cdot {\bf r}_{l}} c_{{\bf k} \beta \alpha \sigma}^{\dagger}$, obtaining $H_{\rm int}^{HF}=\sum_{{\bf k}}H_{\rm int}^{HF}({\bf k})$, where
\begin{equation}\label{HFHybrid}
H_{\rm int}^{HF}({\bf k}) = U\sum_{\beta \alpha \sigma} \langle n_{\beta \alpha \bar{\sigma}}\rangle  n_{ {\bf k} \beta \alpha {\sigma}} - U\sum_{\beta \alpha \sigma} \left(\langle S_{\beta \alpha}^{+}\rangle S_{{\bf k} \beta \alpha}^{-}  + {\rm h.c.}\right).
%\hspace{22.8em}
\end{equation}
The total Hamiltonian under Hartree-Fock approximation can be written as
\begin{equation}\label{ModelHF}
H^{HF} = \sum_{ {\bf k}} \left[H_{KE}({\bf k}) + H_{\rm int}^{HF}({\bf k}) - {1\over 2}\langle H_{\rm int}^{HF}({\bf k}) \rangle\right],
%\hspace{26.8em}
\end{equation}
where the kinetic energy $H_{KE}({\bf k })=\sum_{ \substack{\beta, \alpha, \sigma, \beta{'},\alpha{'}} } \epsilon^{\beta \alpha }_{\beta{'} \alpha{'}} ({\bf k}) c^{\dagger}_{ {\bf k}\beta\alpha \sigma} c_{ {\bf k}\beta{'}\alpha{'}\sigma}$, the matrix element $\epsilon^{\beta \alpha}_{\beta{'} \alpha{'}} ({\bf k}) = \sum_{j} t^{j\beta\alpha}_{0 \alpha{'}\beta{'}} e^{\iota {\bf k} \cdot {\bf r}_{j}} $, and $t^{j\beta\alpha}_{0 \alpha{'}\beta{'}}$ is the hopping amplitude. 

A self consistent calculation in small magnetic unit cells reduces the number of order parameters to converge and also reduces the size of the quadratic Hamiltonian matrices required to diagonalize, leading to relatively faster calculations. For all the parameter points in the phase diagrams shown in the main text, we performed self-consistent calculations for the system of $12 \times 12$ sites
and targeted multiple magnetic unit cell sizes, namely $1\times 1$, $2\times 2$, $4 \times 4$, $6 \times 6$, and $12 \times 12$. We used 10 random initial conditions of the order parameters, for each magnetic unit cell calculation, and declared converged solution with the lowest energy as the ground state. It should be noted that calculation with the magnetic unit cell size of $12 \times 12$ is equivalent to the canonical fully unrestricted Hartree-Fock approach. We found, except for the ALM+FM region with inhomogeneous states, the ground state always lies in the magnetic unit cell sizes smaller than $12\times 12$. To attain self-consistency, we targeted the convergence error of $10^{-8}$ in order parameters. The Anderson mixing method was used to gain accelerated self-consistency.
 
\section{Non-coplanar state in $n$=2 and $n$=4 mean field phase diagrams}
In $n=2$ and 4 phase diagrams, shown in the Fig.~\ref{fig2}(a) and Fig.~\ref{fig3}(b) of the main paper, we found the regions with non-coplanar magnetically ordered ground states. In both cases, we found an 8-sublattice non-coplanar state, namely a state with 8 non-coplanar spin vectors. We found that all the spins on the sublattice B(C), denoted by ${\textrm{B}}_{i}(\textrm{C}_{i})$, lie in the same plane, see Fig.~\ref{Sfig1}(a,b). The angle between the $\textrm{B}_{1} (\textrm{C}_{1})$ and $\textrm{B}_{2} (\textrm{C}_{2})$, and the angle between the two planes with spins $B_{i}$ and $C_{i}$ depends on the location in the phase diagram. However, spins ${\textrm{B}}_{i} (\textrm{C}_{i})$ and $\overline{\textrm{{B}}}_{i} (\overline{\textrm{{C}}}_{i})$  are always aligned opposite to each other. The real-space location of these spins in shown in Fig.~\ref{Sfig1}(b). Interestingly, the sublattice B(C) always has collinear antiferromagnetic one-dimensional chains along $x$ ($y$)-direction, whereas the relative angle between the spins of these chains leads to non-coplanarity. We noticed that strong quasi-one dimensional collinear antiferromagnetic correlations in the above Hartree-Fock results survive even in our exact calculations leading to either a valence bond solid state (on a small Lieb lattice cluster) or effectively decoupled 1D Heisenberg chains (in an effective spin-1/2 model).  
\begin{figure}[!t]
\hspace*{0.0cm}
\vspace*{0cm}
\begin{overpic}[width=0.6\columnwidth]{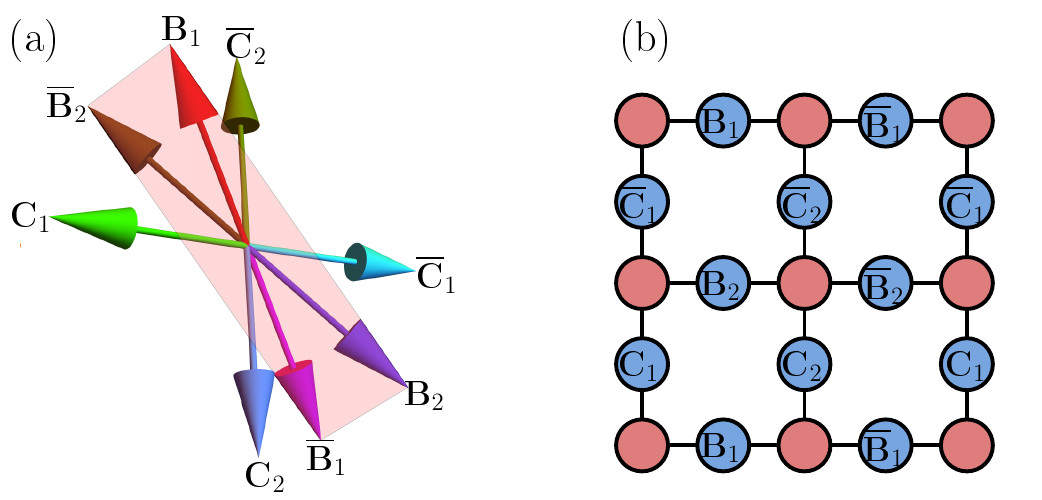}
\end{overpic}
\caption{Panel (a) depicts the 8 spin vectors contributing to the 8-sublattice Non-coplanar state. The positions of these 8 spins on the Lieb lattice is shown in the panel (b).}
\label{Sfig1}
\end{figure}

\section{Single particle spectral function and dynamical structure factor calculations in Exact diagonalization}
We calculated the spin-resolved hole part of the single-particle spectral function using $A_{\sigma}^{h}({\bf q},\omega) = \sum_{\alpha} A_{\alpha\sigma}^{h}({\bf q},\omega)$ and 
\begin{equation}
A_{\alpha\sigma}^{h}({\bf q},\omega) = -\frac{1}{\pi}{\textrm{Im}} [\langle gs|c_{{\bf q}\alpha\sigma}^{\dagger} \frac{1}{\omega+\iota \eta + H - E_{gs}} c_{{\bf q}\alpha\sigma}|gs\rangle]
%\hspace{26.8em}
\end{equation}
 where $c_{{\bf q}\alpha\sigma}^{\dagger}=\frac{1}{\sqrt{L_{x}L_{y}}}\sum_{j}e^{\iota {\bf q}\cdot {\bf r}_{j}} c_{{j}\alpha\sigma}^{\dagger}$, $\alpha$ is the sublattice, and $j$ is the unit cell index.
 
 We also calculated the subalttice resolved dynamical structure factor $S_{\alpha}({\bf q}, \omega)$ defined as
 \begin{equation}
 S_{\alpha}({\bf q}, \omega)=-\frac{1}{\pi}{\textrm{Im}}[\langle gs| S^{+}_{{\bf q}\alpha} \frac{1}{\omega+\iota \eta - H + E_{gs}} S^{-}_{{\bf q}\alpha}|gs \rangle]\\
 %\hspace{26.8em}
 \end{equation}
 where, $S^{+}_{{\bf q}\alpha} = \frac{1}{\sqrt{L_{x}L_{y}}}\sum_{j}e^{\iota {\bf q}\cdot {\bf r}_{j}} S^{+}_{{j}\alpha}$
 . We used the continued fraction approach~\cite{Gagliano01} implemented within Lanczos algorithm for calculating both of the above dynamical response functions. 

\section{Spin-1/2 model on checkerboard lattice: Exact diagonalization results and Linear spin wave theory}
\begin{figure}[H]
\hspace*{3.2cm}
\vspace*{0cm}
\begin{overpic}[width=0.6\columnwidth]{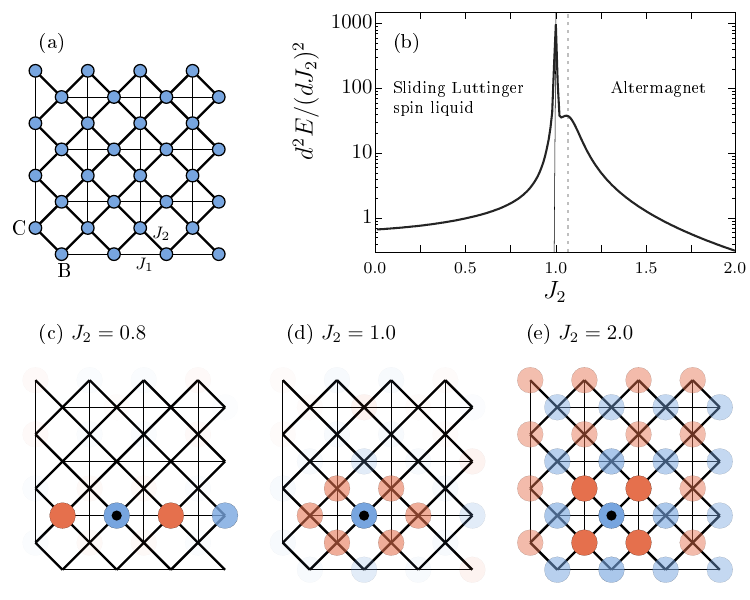}
\end{overpic}
\caption{Panel (a) depicts the 32-site cluster used in ED. The $d^{2}E/(d J_{2})^{2}$ as a function of $J_{2}$ is shown in panel (b). Panels (c), (d), and (e) shows the spin-spin correlation with respect to a site comprising a black circle for $J_{2}$=0.8, 1.0, and 2.0, respectively.}
\label{Sfig2}
\end{figure}
W can describe the low energy properties of the altermagnetic (ALM) Mott state using a spin-${1\over 2}$ Heisenberg model Eq.\ (2) in the main text on a checkerboard lattice depicted in Fig.~\ref{Sfig2}(a), with exchange couplings $J_{1}$ and $J_{2}$. For $n=4$ we estimate  $J_{1} \propto \mathcal{O}(t_{1}^{4}/U^{3})$ and $J_{2} = J_{1} + 4t_{1}^{2}/U + \mathcal{O}(t_{1}^{2}t_{2}/U^{2})$. Similarly for $n=2$, $J_{1} \propto \mathcal{O}(t_{1}^{4}/\epsilon_{A}^{2}U) + \mathcal{O}(t_{1}^{4}/\epsilon_{A}^{3})$ and $J_{2} = J_{1} + 4t_{2}^{2}/U + \mathcal{O}(t_{1}^{2}t_{2}/\epsilon_{A}U) + \mathcal{O}(t_{1}^{2}t_{2}/\epsilon_{A}^{2})$. Using the above estimates, it can be concluded that $J_{2}$ is always larger than $J_{1}$ for $t_{2}/t_{1}>0$. For $t_{2}/t_{1}<0$, the third-order terms are negative. For $n=4$, the magnitude of the third order term is always smaller than the second order term in the limit $U\gg|t_{1}|, |t_{2}|$, so we again expect $J_{2}>J_{1}$. However, for $n=2$, in the limit $U\gg\epsilon_{A}>|t_{1}|,|t_{2}|$ third order term can be large which may lead to $J_{2}<J_{1}$. The above heuristic arguments suggest that the $J_{2}/J_{1}>1$ is most plausible in the strong coupling limit. Nevertheless, in our study we have solved the spin model for arbitrary $J_{2}/J_{1}>0$ to investigate the possible ground states.

The $32-$site cluster with PBC shown in Fig.~\ref{Sfig2}(a) is solved using exact diagonalization Lanczos algorithm. We calculated the $d^{2}E/(d J_{2})^{2}$ as a function of $J_{2}$ for fixed $J_{1}=1.0$, see Fig.~\ref{Sfig2}(b). The large peak present near $J_{2}=1.0$ suggests a first-order phase transition. For $J_{2}<1.0$, we noticed that the system effectively decouples into one-dimensional chains with antiferromagnetic correlations along the $x(y)$ direction for the sublattice $B(C)$. The spin-spin correlation with respect to a site belonging to sublattice B (shown with a black circle in the center) is shown in Fig.~\ref{Sfig2}(c). It is expected that for large systems the spin-spin correlations will decay algebraically following the Luttinger liquid behavior as in 1D antiferromagnetic Heisenberg chains, hence called sliding Luttinger spin liquid. Above the first-order phase transition in the region $1 \leq J_{2}\lesssim1.07$, the system shows exponentially decaying spin-spin correlations as depicted in Fig.~\ref{Sfig2}(d). The system quickly grows antiferromagnetic correlations as $J_{2}$ is further increased, accompanied by a small peak in $d^{2}E/(d J_{2})^{2}$ near $J_{2}=1.07$, Fig.~\ref{Sfig2}(b). We show the spin-spin correlation for $J_{2}=2.0$ in Fig~\ref{Sfig2}(e), depicting system-spanning antiferromagnetic correlations. Our main finding, that for the large range of $J_{2}>J_{1}$ the ground state has staggered antiferromagnetic correlations, is consistent with multiple numerical studies performed on the checkerboard lattice. 

In addition to calculating exact dynamical spin structure factor, we also used linear spin wave theory (LSWT) and found two magnon modes with dispersion 
\begin{equation}
\omega^{\pm}({\bf q})= {1\over 4}\left(\pm (\lambda_{q_{x}}-\lambda_{q_{y}})+\sqrt{ ( \lambda_{q_{x}}+\lambda_{q_{y}})^{2} - 4|\xi_{\bf q}|^{2} }\right),
\end{equation}
where $\lambda_{q_{x(y)}}=4J_{2} + 2J_{1}(\cos{q_{x(y)}}-1)$, and $\xi_{\bf q}=4J_{2}\cos(q_{x}/2)\cos(q_{y}/2)$. The LSWT theory results are depicted in Fig.~\ref{fig5} of the main paper and show qualitatively similar results to exact diagonalization. There are two differences between the LSWT and exact results, (i) the LSWT magnons have to be normalized by a factor of $1.35$ to obtain the correct magnon bandwidth, and (ii) LSWT does not capture minima at point $M$ along $M$ to $X(Y)$ direction in $S_{B(C)}({\bf q},\omega)$. In the exact calculation, we also noticed signatures of deconfined spinons at high energies which will be interesting to explore in future studies.

%Using LSWT (Holstein-Primakoff), discuss splitting in chiral magnons.

\section{Electronic bands in altermagnetic metals}
\begin{figure}[H]
\hspace*{0.0cm}
\vspace*{0cm}
\begin{overpic}[width=1.0\columnwidth]{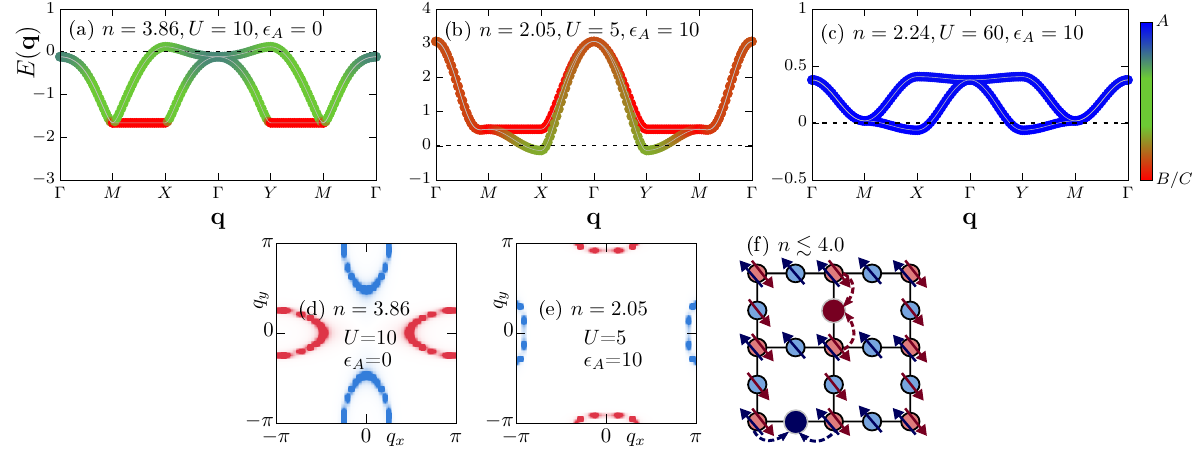}
\end{overpic}
\caption{Panels (a,b,c) show the sublattice resolved single-particle bands for the representative parameter points written inside the respective panels, the color indicates the contribution from the sublattices A and B/C (B or C). The spin-resolved Fermi surfaces for the parameter points $(n=3.86,U=10, \epsilon_{A}=0)$ and $(n=2.05, U=5, \epsilon_{A}=10)$ are shown in panels (d) and (e), respectively. Panel (f) depicts the real space kinetic motion of the doped holes in $n \lesssim 4$ ALM metal.}
\label{Sfig3}
\end{figure}
In the main text, we established the presence of ALM metallic regions by doping the Mott state. In this section we provide more details about the ALM metallic phases by discussing the sublattices' contribution at the Fermi surfaces of these metals. For the  $n\lesssim 4$ ALM metal, we find that the Fermi surface, shown in Fig.~\ref{Sfig3}(d), constitutes cigar-shaped spin-up and down hole-pockets near $Y$ and $X$ points, respectively, with considerable contribution from all three sublattices, Fig.~\ref{Sfig3}(a). The cigar-shape of hole-pockets can be explained by the constrained motion of spin  up (down)electrons along $x$ ($y$) direction, depicted in Fig.~\ref{Sfig3}(f), which results from the Pauli blocking of the spin up (down) electron motion in $y$ ($x$) direction due to doubly occupied sublattice A. However, the diagonal hopping $t_{2}$ renormalizes the above quasi-1D picture and also helps to stabilize FM polarons above some critical value of $U$. 

The band structure of ALM metal in the $n\gtrsim 2$, $U<\epsilon_{A}$ region is shown in Fig.~\ref{Sfig3}(b). We again find a considerable contribution from all three sublattices at the Fermi surface accompanied by the $d$-wave spin-spliting, Fig.\ref{Sfig3}(e). Interestingly, in comparison to the other ALM metallic phases,  Fig.~\ref{Sfig3}(b) indicates much smaller band splitting along the momentum cuts $M \rightarrow X \rightarrow \Gamma$ and $\Gamma \rightarrow Y\rightarrow M$. This small splitting suggests that a weaker higher-order process is the leading cause of the anisotropy in the effective hopping for doped electrons. 

Finally, we also show the band-structure for the ($n\gtrsim 2$, $U>\epsilon_{A}$) ALM metal in the Fig.~\ref{Sfig3}(c). We notice that the electronic bands near the Fermi-surface are mostly contributed by the sublattice A, which is consistent with the real-space picture provided in the main-paper. Moreover, we confirmed that the energy difference between the $\Gamma$ and $X$ ($Y$) momentum points also matches the expected value of $4{t_{\rm eff}}$, bandwidth of a 1D tight-binding model.

 \section{Inhomogeneous states in doped Mott insulators}
 \begin{figure}[!t]
\hspace*{0cm}
\vspace*{0cm}
\begin{overpic}[width=0.6\columnwidth]{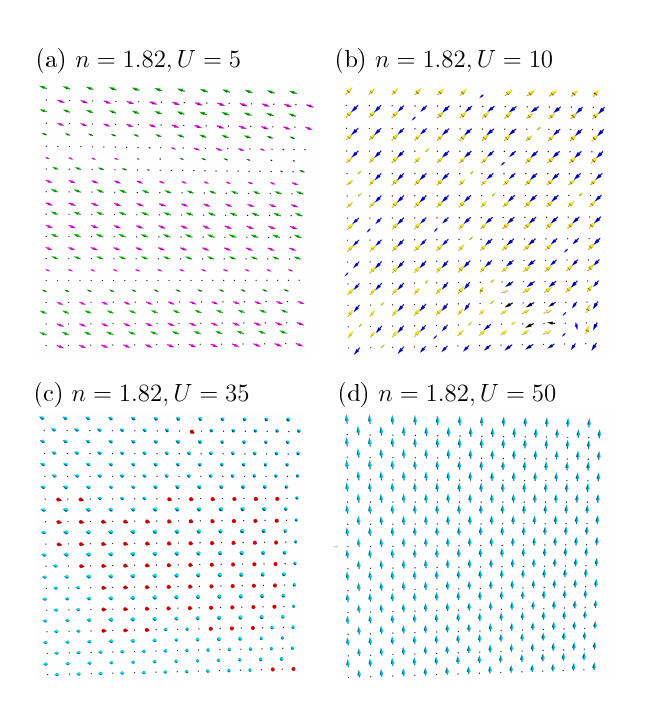}
\end{overpic}
\caption{Above panels show the spins in the real space attained by the unrestricted Hartree-Fock, for the respective parameter values. The direction and color of the arrows depict the direction of the spins ${\hat{S}}_{j\alpha}$, and the length of the arrows represent the magnitude of the spins $|{\bf S}_{j\alpha}|$.}
\label{Sfig4}
\end{figure}
Substantial parts of the doping $n$ vs.\ $U$ phase diagrams, seen in the  hole/electron-doped $n=4$ and hole-doped $n=2$ regions of Fig.~\ref{fig6}(a,b) of the main paper, feature inhomogeneous states with the combination of ALM and ferromagnetic (FM) patches. For $n<2$ ($n>4$) the doped holes (electrons) mostly reside on sublattices B and C because in the undoped case sublattice A is already nearly empty (fully occupied). As a result, the doped fermions effectively move on an antiferromagnetic bipartite square lattice via $t_{2}$ hoppings, hence an instability towards ferromagnetism is expected. In this section, we illustrate the evolution of magnetism in the doped state as $U$ is increased, using the hole-doped $n=2$ case with $n=1.82$ as an example. Similar results were observed in all the region of the phase diagrams dubbed ``ALM+FM". In the limit of weak $U$, see Fig.~\ref{Sfig4}(a), the system shows phase separation: in one region doped holes are accumulated quenching the spin moments, while the other region has nearly half-filled sublattices B and C with the ALM order intact. We found that the region where doped holes are present quickly develops FM order as $U$ is increased, leading to an inhomogeneous state with FM polarons in the background of the ALM state, see Fig.~\ref{Sfig4}(b). On further increasing $U$, see Fig.~\ref{Sfig4}(c), the FM polarons merge to give rise to a state with phase separation between the ALM and FM domains. These ALM+FM inhomogeneous states are precursor of the fully polarized FM state present in the large $U$ limit, which we also observed as depicted in the Fig.~\ref{Sfig4}(d). In phase diagrams shown in the main paper, for brevity, we use the expression ``ALM+FM" for all the regions where the above features are observed. 

%
%\usepackage{titlesec}
%\titleformat{\section}{\large\bfseries}{\thesection}{5em}{}

%    \bibliographystyle{apalike} % bibliography style - recommend using apalike-doi as it hyperlinks DOIs   

%-----------------------------------------------------------------

}
%\usepackage{titlesec}
%\titleformat{\section}{\large\bfseries}{\thesection}{5em}{}


\begin{thebibliography}{10}
%\bibliographystyle{ieeetr} 

%the first 4
\bibitem{Hayami01} S. Hayami, Y. Yanagi, and H. Kusunose, {\href{https://journals.jps.jp/doi/10.7566/JPSJ.88.123702}{J. Phys. Soc. Jpn. {\bf 88}, 123702 (2019)}}.

\bibitem{Smejkal03} L. {\v{S}}mejkal, R. Gonz\'alez-Hern\'andez, T. Jungwirth, and
J. Sinova, {\href{https://www.science.org/doi/10.1126/sciadv.aaz8809}{Sci. Adv. {\bf 6}, eaaz8809 (2020)}}.

\bibitem{Yuan01} L.-D. Yuan, Z. Wang, J.-W. Luo, E. I. Rashba, and A. Zunger, {\href{https://journals.aps.org/prb/abstract/10.1103/PhysRevB.102.014422}{Phys. Rev. B {\bf 102}, 014422 (2020)}}.

\bibitem{Mazin01} I. I. Mazin, K. Koepernik, M. D. Johannes, R. Gonz\'alez-Hern\'andez, and L. {\v{S}}mejkal , {\href{https://www.pnas.org/doi/full/10.1073/pnas.2108924118}{Proc. Natl. Acad. Sci. U.S.A. {\bf 118}, e2108924118 (2021)}}.


\bibitem{Smejkal01} L. {\v{S}}mejkal, J. Sinova, and T. Jungwirth, {\href{https://journals.aps.org/prx/abstract/10.1103/PhysRevX.12.031042}{Phys. Rev. X {\bf 12}, 031042 (2022)}}.

\bibitem{Smejkal02} L. {\v{S}}mejkal, J. Sinova, and T. Jungwirth, {\href{https://journals.aps.org/prx/abstract/10.1103/PhysRevX.12.040501}{Phys. Rev. X {\bf 12}, 040501 (2022)}}.

%topology using alm
\bibitem{YLi01} Y.-X. Li and C.-C. Liu, {\href{https://journals.aps.org/prb/abstract/10.1103/PhysRevB.108.205410}{Phys. Rev. B {\bf 108}, 205410 (2023)}}.
\bibitem{Zhu01} D. Zhu, Z.-Y. Zhuang, Z. Wu, and Z. Yan, {\href{https://journals.aps.org/prb/abstract/10.1103/PhysRevB.108.184505}{Phys. Rev. B {\bf 108}, 184505 (2023)}}.
\bibitem{Heung01} T. F. Heung and M. Franz, {\href{https://arxiv.org/abs/2411.17964}{arXiv:2411.17964}}.



%CrSb
\bibitem{Ding01} J. Ding et al.,{\href{https://journals.aps.org/prl/abstract/10.1103/PhysRevLett.133.206401}{Phys. Rev. Lett. {\bf 133}, 206401 (2024)}}.
\bibitem{Lu01} W. Lu et al., {\href{https://arxiv.org/abs/2407.13497}{arXiv:2407.13497}}.
\bibitem{CLi01} C. Li et al., {\href{https://arxiv.org/abs/2405.14777}{arXiv:2405.14777}}.


%MnTe
\bibitem{JKrempasky01} J. Krempask\'y et al., {\href{https://www.nature.com/articles/s41586-023-06907-7}{Nature {\bf 626}, 517-522 (2024)}}.
\bibitem{Lee01} S. Lee et al., {\href{https://journals.aps.org/prl/abstract/10.1103/PhysRevLett.132.036702}{Phys. Rev. Lett. {\bf 132}, 036702 (2024)}}.
\bibitem{Osumi01} T. Osumi et al., {\href{https://journals.aps.org/prb/abstract/10.1103/PhysRevB.109.115102}{Phys. Rev. B {\bf 109}, 115102 (2024)}}.

\bibitem{Liu01} J. Liu et al., {\href{https://journals.aps.org/prl/abstract/10.1103/PhysRevLett.133.176401}{Phys. Rev. Lett. {\bf 133}, 176401 (2024)}}.
\bibitem{Kessler01} P. Ke{\ss}ler et al., {\href{https://www.nature.com/articles/s44306-024-00055-y}{npj Spintronics {\bf 2}, Article number: 50 (2024)}}.



%FeSe
\bibitem{Mazin02} I. Mazin, R.-Gonz\'alez-Hern\'andez, L. {\v{S}}mejkal, {\href{https://arxiv.org/abs/2309.02355}{arXiv:2309.02355 (2023)}}.

%La2O3Mn2Se2
\bibitem{Ni01} N. Ni, E. Climent-Pascual, S. Jia, Q. Huang, and R. J. Cava, {\href{https://journals.aps.org/prb/abstract/10.1103/PhysRevB.82.214419}{Phys. Rev. B {\bf 82}, 214419 (2010)}}.

\bibitem{Wei01} C.-C. Wei et al., {\href{https://arxiv.org/abs/2410.14542}{arXiv:2410.14542}}.

%La2Co2O3Se2 (Vortex)
\bibitem{Fuwa01} Y. Fuwa, T. Endo, M. Wakeshima, Y. Hinatsu, K. Ohoyama, {\href{https://pubs.acs.org/doi/10.1021/ja109007g}{J. Am. Chem. Soc., {\bf 132}, 18020-18022 (2010)}}.

%La2Fe2O3S(Se)2 : 2k
\bibitem{Freelon01} B. Freelon et al., {\href{https://journals.aps.org/prb/abstract/10.1103/PhysRevB.99.024109}{Phys. Rev. B {\bf 99}, 024109 (2019)}}.
%Sr2F2Fe2OS2 : 2k
\bibitem{LZhao01} L. L. Zhao, S. Wu, J. K. Wang, J. P. Hodges, C. Broholm, and E. Morosan, {\href{https://journals.aps.org/prb/abstract/10.1103/PhysRevB.87.020406}{Phys. Rev. B {\bf 87}, 020406(R) (2013)}}.


%KV2Se2O
\bibitem{BJiang01} B. Jiang et al., {\href{https://arxiv.org/abs/2408.00320}{arXiv:2408.00320}}.
%RbV2Te2O
\bibitem{FZhang01} F. Zhang et al., {\href{https://arxiv.org/abs/2407.19555}{arXiv:2407.19555}}.


%Titanium based oxypnictides
%https://iopscience.iop.org/article/10.1088/0953-8984/25/36/365501
%https://pubs.acs.org/doi/epdf/10.1021/cm010009f?ref=article_openPDF
%https://www.mdpi.com/2410-3896/2/1/4
%https://journals.jps.jp/doi/abs/10.1143/JPSJ.81.103706


%V2Se2O monolayer , theoretically proposed 2d altermagnet
%https://www.nature.com/articles/s41467-021-23127-7



\bibitem{Brekke01} B. Brekke, A. Brataas, and A. Sudb{{\o}}, {\href{https://journals.aps.org/prb/abstract/10.1103/PhysRevB.108.224421}{Phys. Rev. B {\bf 108}, 224421 (2023)}}.

\bibitem{Antonenko01} D. S. Antonenko, R. M. Fernandes, and J. W. F. Venderbos , {\href{https://arxiv.org/abs/2402.10201}{arXiv:2402.10201}}.


%something important
%https://arxiv.org/pdf/2402.18629


%LSWT underestimates the magnon velocity
%https://journals.aps.org/prb/pdf/10.1103/PhysRevB.79.195102
%https://journals.aps.org/prb/pdf/10.1103/PhysRevB.92.195145

%Fractionaliations in high energy spectra, well known phenomenon
%https://www.nature.com/articles/nphys3172


%Hubbard model studies
\bibitem{Maier01} T. A. Maier and S. Okamoto, {\href{https://journals.aps.org/prb/abstract/10.1103/PhysRevB.108.L100402}{Phys. Rev. B {\bf 108}, L100402 (2023)}}.
\bibitem{Leeb01} V. Leeb, A. Mook, L. {\v{S}}mejkal, and J. Knolle, {\href{https://journals.aps.org/prl/abstract/10.1103/PhysRevLett.132.236701}{Physl Rev. Lett. {\bf 132}, 236701 (2024)}}.
\bibitem{Das01} P. Das, V. Leeb, J. Knolle, and M. Knap, {\href{https://journals.aps.org/prl/abstract/10.1103/PhysRevLett.132.263402}{Phys. Rev. Lett {\bf 132}, 263402(2024)}} .
\bibitem{Roig01} M. Roig, A. Kreisel, Y. Yu, B. M. Andersen, and D. F. Agterberg, {\href{https://journals.aps.org/prb/abstract/10.1103/PhysRevB.110.144412}{ Phys. Rev. B {\bf 110}, 144412 (2024) }}.
\bibitem{Ferrari01} F. Ferrari and R. Valenti, {\href{https://journals.aps.org/prb/abstract/10.1103/PhysRevB.110.205140}{Phys. Rev. B {\bf 110}, 205140 (2024) }}.
\bibitem{Re01} L. D. Re, {\href{https://arxiv.org/abs/2408.14288}{arXiv:2408.14288}}.
\bibitem{Giuli01} S. Giuli, C. M.-Zaera, and M. Capone, {\href{https://arxiv.org/abs/2410.00909}{arXiv:2410.00909}}.



%standard Lieb lattice H-F
\bibitem{Gouveia01} J. D. Gouveia and R. Dias, {\href{https://www.sciencedirect.com/science/article/abs/pii/S0304885315001353}{J. Magn. Magn. Mater. {\bf 382}, 312 (2015)}}.

%Checkerboard Spin=1/2 literature
\bibitem{Sindzingre01} P. Sindzingre, J.-B. Fouet, and C. Lhuillier, {\href{https://journals.aps.org/prb/abstract/10.1103/PhysRevB.66.174424}{Phys. Rev. B {\bf 66}, 174424 (2002)}}.
\bibitem{YHChan01} Y.-H. Chan, Y.-J. Han, and L.-M. Duan, {\href{https://journals.aps.org/prb/abstract/10.1103/PhysRevB.84.224407}{Phys. Rev. B {\bf 84}, 224407 (2011)}}.
\bibitem{SMoukari01} S. Moukouri, {\href{https://journals.aps.org/prb/abstract/10.1103/PhysRevB.77.052408}{Phys. Rev. B {\bf 77}, 052408 (2008)}}.
\bibitem{RFBishop01} R. F. Bishop, P. H. Y. Li, D. J. J. Farnell, J. Richter, and C. E. Campbell, {\href{https://journals.aps.org/prb/abstract/10.1103/PhysRevB.85.205122}{Phys. Rev. B {\bf 85}, 205122 (2012)}}.


\bibitem{Supple} Supplementary link{\href{}{}}.

\bibitem{PIgoshev01} P. A. Igoshev, M. A. Timirgazin, A. A. Katanin, A. K. Arzhnikov, and V. Yu. Irkhin, 
{\href{https://journals.aps.org/prb/abstract/10.1103/PhysRevB.81.094407}{Phys. Rev. B {\bf 81}, 094407 (2010)}}.

\bibitem{JXu01} J. Xu, C.-C. Chang, E. J Walter and S. Zhang, {\href{https://iopscience.iop.org/article/10.1088/0953-8984/23/50/505601}{J. Phys. Condens. Matter {\bf 23}, 505601 (2011)}}.

\bibitem{RSamajdar01} R. Samajdar and R. N. Bhatt, {\href{https://journals.aps.org/prb/abstract/10.1103/PhysRevB.109.235128}{Phys. Rev. B {\bf 109}, 235128 (2024)}}.

\bibitem{RSamajdar02} R. Samajdar and R. N. Bhatt, {\href{https://journals.aps.org/pra/abstract/10.1103/PhysRevA.110.L021303}{Phys. Rev. A {\bf 110}, L021303 (2024)}}.

\bibitem{THanisch01} T. Hanisch, B. Kleine, A. Ritzl, E. M.-Hartmann, {\href{https://onlinelibrary.wiley.com/doi/abs/10.1002/andp.19955070405}{Ann. Phys. {\bf 507}, 303–328 (1995)}}.
\bibitem{VBach01} V. Bach, E. H. Lieb, M. V. Travaglia, {\href{https://www.worldscientific.com/doi/abs/10.1142/S0129055X06002735}{Rev. Math. Phys. {\bf 18}, 519–543 (2006)}}.
\bibitem{BShastry01} B. S. Shastry, H. R. Krishnamurthy, and P. W. Anderson, {\href{https://journals.aps.org/prb/pdf/10.1103/PhysRevB.41.2375}{Phys. Rev. B {\bf 41}, 2375 (1990)}}.


\bibitem{TYajima01} T. Yajima, K. Nakano, F. Takeiri, T. Ono, Y. Hosokoshi, Y. Matsushita, J. Hester, Y. Kobayashi, and H. Kageyama, {\href{https://journals.jps.jp/doi/abs/10.1143/JPSJ.81.103706}{J. Phys. Soc. Jpn. {\bf 81}, 103706 (2012)}}.

\bibitem{XYan01} X.-W. Yan and Z.-Y. Lu {\href{https://iopscience.iop.org/article/10.1088/0953-8984/25/36/365501}{J. Phys. Condens. Matter {\bf 25}, 365501 (2013)}}.

\bibitem{Dagotto01} E. Dagotto, {\href{https://journals.aps.org/rmp/abstract/10.1103/RevModPhys.66.763}{Rev. Mod. Phys. {\bf 66}, 763 (1994)}}.

\bibitem{MDurrnagel01} M. D\"urrnagel, H. Hohmann, A. Maity, J. Seufert, M. Klett, L. Klebl, and R. Thomale, {\href{https://arxiv.org/pdf/2412.14251}{arXiv:2412.14251}.}

%\bibitem{} {\href{}{}}.
\end{thebibliography}

\begin{thebibliography}{10}
%\bibliographystyle{ieeetr} 


\bibitem{Gagliano01} E. R. Gagliano and C. A. Balseiro, {\href{https://journals.aps.org/prl/abstract/10.1103/PhysRevLett.59.2999}{ Phys. Rev. Lett. {\bf 59}, 2999 (1987)}}.

%\bibitem{} {\href{}{}}.
\end{thebibliography}
\end{document}